# Ultraprecise Detection of Low Molecular Weight Antibodies Using Mid-IR Toroidal Plasmonic Meta-Atoms

Burak Gerislioglu,[1][†] Zeinab Ramezani,[2] S. Amir Ghoreishi,[3] and Arash Ahmadivand[4][*][†]

[1]*Department of Physics and Astronomy, Rice University, 6100 Main St, Houston, Texas 77005, United States*

[2]*Electrical and Computer Engineering Department, Semnan University, Semnan, Iran*

[3]*Faculty of Electrical and Computer Engineering, Science and Research Branch, Islamic Azad University of Tehran, Tehran, Iran*

[*]aahmadiv@rice.edu

†: Equal contribution

**Abstract:**

Plasmonic metamaterials can be used to enhance mid-infrared sensors and detectors for immunosensing applications. The major challenge is to integrate these technologies in low-cost, compact, and promising platforms. In this study, as an emerging approach, we utilized toroidal resonant plasmonic metasurfaces to develop an ultrasensitive label-free analytical platform for the detection of specific antibiotics in the mid-IR regime. Taking advantage of unique sensitivity of robustly squeezed electromagnetic fields in toroidal meta-atoms, we validated that our proposed technique based on toroidal metastructures can provide sensing of biological targets with the weight of ~0.6 KDa at attomolar concentrations. We believe that this understanding extends the capabilities of plasmonic metamaterials analyze the presence of specific biological targets with high precision.

Plasmonic metachips have revitalized advanced and sensitive label-free biomarker detection techniques by enabling robust electromagnetic field confinement down to extreme-subwavelength scales, which enabled the detection of biomolecules at ultralow densities with high precision.[1,2] This counterpart of nanophotonics have promoted the development of cost-effective multi-analyte biosensors, tailored for ultrafast and real-time recognition of diverse infection and diseases fingerprints.[1-3] Considering the hype-cycle profile of plasmonic biosensors technology, currently, these devices are well into a stable commercialization phase.[4] Plasmonic biosensing devices majorly operate based on resonant nanostructures, are very sensitive to the dielectric materials in contact with the surface of the nanochips. While visible/near-infrared wavelength-resonant structures are of particular interest for optic-based pharmacological applications, the mid-infrared (mid-IR) bandwidth is particularly well-matched for nanometric biomolecule sensorics.[5,6] In general, the mid-IR spectrum encircles the vibrations of most of the biological objects including envelope proteins, DNA, and lipids.[7] Although the vibration signals of biomolecules are dramatically weak, as a promising technique in accessing these fingerprints, mid-IR spectroscopy facilitates a non-invasive and non-destructive label-free biosensing approach.[8,9] To this end, resonant plasmonic metamaterials based on artificial building blocks have extensively been employed for mid-IR biosensing by taking advantage of the displacements in the position of the induced moments due to the variations in the dielectric permittivity of media.[8-11] Traditionally, Fano-resonant plasmonic metamaterials have immensely been used for mid-IR detection due to showing substantial sensitivity of the induced asymmetric lineshape to the environmental perturbations.[10,12,13] In spite of the fact that Fano resonant sensors provide accurate detection of low-weight biological targets at low concentrations, this approach does not efficiently address the identification of ultralow-weight proteins (i.e. beta-amyloid, tau protein) and viruses (i.e. Zika, PRD1, MS2) at early-stage of diseases due to poor limit of detection (LOD) in assays.

Very recently, a new class of label-free, on-site, selective, and extremely-sensitive plasmonic biosensors has been developed and introduced based on toroidal multipole electrodynamics.[14-16] In the field of theoretical physics, optically driven dynamic toroidal moments constitute an independent group of elementary electromagnetic sources with dramatically weak far-field radiation patterns.[17,18] The fundamental member of toroidal multipoles is a toroidal dipole that can be created by the polarization currents flowing on a surface of a torus along its meridian.[17-24] From a more applied perspective, due to possessing a spectrally narrow response and strongly squeezed local field enhancement, a toroidal dipole manifests higher sensitivity in practical applications. This makes toroidal metamaterials as potential alternatives for conventional optical signal transduction mechanisms. This can be perceived by considering the radiated electric field from the classical $E_p = n^2 k_0^2 (\hat{r} \times \hat{r} \times \hat{p}/r)$ and toroidal $E_T = n^3 k_0^3 (\hat{r} \times \hat{r} \times \hat{T}/r)$ scatterers, and their relation to the refractive index ($n$) variations of media.[25] This feature enables quantitative fingerprinting of extreme-subwavelength multimolecular aggregates with remarkable precision that cannot be touched with existing detection systems.

In this work, we demonstrate the attomolar optic-based sensing of ultralow weight (~582.6 Da) antibiotic molecules (kanamycin sulfate; $C_{18}H_{36}N_4O_{11} \times H_2SO_4$) using our developed toroidal mid-IR plasmonic metamaterial. The designed metachip consists of a specifically engineered symmetric multipixel gold nanostructure with an ability to sustain a strong toroidal dipole along the mid-IR spectral range around $\lambda$~5250 nm ($\omega$~1904.7 cm$^{-1}$). The preliminary quantitative studies for the refractive index variations (Supporting Information) verified the unique sensitivity of the system to the minor perturbations in the dielectric properties of the media. As a proof of principle, we detected the presence of kanamycin sulfate molecules attached to the plasmonic metamolecules at attomolar concentrations around 5.2 attomolar (600 molecules in 200 µL).

**The spectral response of the plasmonic mid-IR unit cell**. To begin with, we analyze the spectral response of the devised plasmonic mid-IR unit cell by considering both numerical and experimental studies. Figure 1a illustrates the tailored symmetric plasmonic unit cell, picturing the formation of a spinning closed-loop charge-current configuration around the center of the meta-atom (not to scale). This panel contains the judiciously defined geometries of the unit cell based on finite-difference time-domain (FDTD) numerical analysis (see Methods). A scanning electron microscopy (SEM) image of the fabricated metallic aperture arrays is exhibited in Figure 1b (For fabrication details, see Methods). Considering the mismatch between the induced magnetic fields and subsequently generated displacement currents ($\vec{J}$) in proximal nanoresonators under normalized *y*-polarized beam illumination (Figure S1, Supporting Information), we accurately validated the excitation of a toroidal dipole feature at $\lambda$~5250 nm ($\omega$~1904.7 cm$^{-1}$), as shown in Figure 1c. In this profile, the blue and orange solid lines explicitly shows the numerically and experimentally obtained transmission spectra, respectively, under transverse light illumination. Conversely, for the *x*-polarized beam (longitudinal) beam, the toroidal feature vanishes due to losing the circulating magnetic fields in adjacent resonators (see dashed lines in Figure 1c). To show the influence of the nanosize gaps on the excitation of oppositely rotating magnetic-fields (m) in the unit cell, we depicted the distributions of the electric-field (E-field) in a 3D map in Figure 1d. This profile clearly illustrates the intense confinement of plasmons at the capacitive gap spots, which acting as a key parameter in defining the formation of toroidal moment.[26] Moreover, the presence and accumulation of proteins, enzymes, molecules, etc. at these capacitive openings dramatically changes the spectral response of the metasensor due to affecting the induced magnetic field intensity and toroidal mode spinning properties. The theoretical predictions are confirmed by quantifying the near-field enhancement around the toroidal structure as a function of operating bandwidth. As plotted in Figure 1e, the maximum enhancement of $|E|^2$ averaged over the meta-atom along the *z*-axis ($E_z$) is occurred at the toroidal dipole wavelength, gives rise to large spectrally selective enhancement of the plasmonically boosted near-field and holds strong promise for biosensing applications.

The nature of the optically driven transmission dip and the scenario to explain the observed spinning moment are explained and represented in Supporting Information. In this addendum file and Figure S2, we theoretically calculated the radiated power from the multipoles, verifying the suppression of the electric dipole, and simultaneous superposition of magnetic quadrupole and toroidal dipole around the resonance wavelength. The obtained results are consistent with our computations for the displacement current in Figure S1. Furthermore, as an alternative route to verify the formation of head-to-tail toroidal mode, we numerically calculated and plotted the cross-sectional (*xz*-plane) vectorial magnetic-field map for the spinning charge-current arrangement in Figure S3. The illustrated graph strongly validates the formation of head-to-tail toroidal field across the metastructure.

**Detection of antibiotic molecules by toroidal plasmonic unit cell.** In continue, the detection performance and sensing properties of the developed mid-IR toroidal metamaterial will be explained. To verify the sensitivity of the fabricated metasurface, we qualitatively evaluated the variations in the plasmonic response of the metasensor by varying the refractive index of the medium, as demonstrated in Figure S4 (Supporting Information). In general, the lineshape of a toroidal dipole is very narrow, which allows for a highly accurate measurement of small shifts in resonance wavelength induced by variations in the dielectric permittivity of the unit cell environment.[15,16] Conventionally, the ratio of the plasmonic resonant moment energy shift per refractive index unit perturbation in the surrounding ambience defines the precision of an optical biosensor.[27,28] To investigate the sensitivity of the toroidal-resonant meta-atom, we performed a set of experiments by embedding the mid-IR metamaterial in different dielectric liquids (water, acetone, paraldehyde). Figures S4a and S4b show a giant and continuous red-shift in the position of the toroidal dip by increasing the refractive index of the media ($\Delta\lambda_T$~950 nm ($\Delta\omega_T$~304.7 cm$^{-1}$) for $\Delta n$~0.4). These significant variations in the toroidal dipole wavelength stems from its higher sensitivity to the environmental perturbations. Taking advantage of this feature, we utilize the fabricated plasmonic metasurface for the detection of ultralow weight (~582.6 Da) antibiotic molecules (kanamycin sulfate).

Recognition of the presence of the antibiotic molecules around the toroidal meta-atoms is performed by using the transmission difference between two different regimes (absence and presence of biomolecules) as: $\Delta T(\omega) \equiv \left|t_{yy}^{Water}(\omega)\right|^2 - \left|t_{yy}^{Molecule}(\omega)\right|^2$, in which $t_{yy}$ is the tensor correlating with the transmitted and incident electric fields through the metasurface under *y*-polarized light exposure. In Figure 2a, an artistic picture of the toroidal unit cell in the presence of targeted biomolecule is depicted, showing the kanamycin sulfate molecules on the metadevice (The inset is the molecular structure of kanamycin sulfate). By mixing the kanamycin sulfate molecules with distilled water (see the associated dark- and bright-field images for the diffraction pattern and transmission electron microscope (TEM) images of molecules, shown in Figures 2b and 2c, respectively), we introduced the prepared solution (~10 µL) to the plasmonic metasurface, as shown in the magnified SEM image in Figure 2d. The accumulation of kanamycin sulfate molecules at the capacitive openings dramatically changes the measured transmission spectra, as shown in Figure 2e. Following the same route as the earlier studies in Figure S4, by considering water as the reference point, we observed significant red-shift in the position of the toroidal dipole in the range of 5400 nm<$\lambda_T$<6600 nm (1515.1 cm$^{-1}$<$\omega_T$<1851.8 cm$^{-1}$), while the concentration of targeted molecules continuously increases from 0.1 fmol to 10 fmol. Figure 2f exhibits the toroidal dipole shift as a function of the concentration of kanamycin sulfate molecules, where the slope of the red-shift of the asymmetric lineshape is significantly sharp from 0.1 fmol to 1 fmol concentrations (Green region), validates the LOD of around ~0.85 fmol (850 amol) for the metasensor. This trend was observed for much denser concentrations of kanamycin sulfate molecules with moderate slope (Red region) due to destructive influence on the formation of toroidal moment, which cancels the mismatch between the rotating magnetic fields and induced surface current densities. To investigate the origin of the resonance shift due to the presence of biomolecules, one should apply both the permittivity variations and near-field coupling in the corresponding computations:[29]

$$\frac{\Delta \lambda_T}{\lambda_T} = -\frac{1}{2} \left( \frac{\int_0^t \mathbf{E}_T(\mathbf{r}).(\hat{\varepsilon}-1).\mathbf{E}_T(\mathbf{r}) d\mathbf{r}}{\int_0^\infty |\mathbf{E}_T(\mathbf{r})|^2 d\mathbf{r}} \right) \quad (1)$$

where $t$ is the thickness of the dropped liquid layer on the surface of the metachip, $\hat{\varepsilon}$ is the permittivity tensor, and $|\mathbf{E}_T(\mathbf{r})|$ is the near-field at the gaps. Technically, $\Delta \lambda_T$ possesses direct influence on the dephasing time and resonant wavelength position of the toroidal metasensor. To exhibit the notion of structure-specific biological detection,[30] we considered the dependency of the spectral peak on the wavelength mismatch between the toroidal resonance moment wavelength and the biological vibrational modes including the dephasing time of modes, the dipolar mode strength, and bond orientations. Figure 3a illustrates the impact on the induced toroidal resonance lineshape and the corresponding dephasing time due to the modified transmission tensor. This panel indicates the dephasing time as a function of antibiotic concentration, showing the decay in the toroidal resonance lifetime for both numerically (shaded curve) and experimentally (circles) recorded lineshapes (computed by $\tau = 2\hbar/\text{FWHM}$).[31] Although by increasing the concentration of molecules, the toroidal dipole continuously and monotonically decays, the induced moment keeps its quality reasonably for higher concentrations. In Figure 3b, we plotted the experimental results for the before and after binding of the antibiotic to the metallic meta-atoms, indicating the variations in the transmissivity ratios ($\Delta T/T$) of the toroidal moment.

**Conclusions**

In conclusion, we have demonstrated a novel application of toroidal resonant plasmonic metamaterial to detect ultralow-weight antibiotic molecules at attomolar concentrations along the mid-IR spectra. As a proof of principle, we obtained an exquisite detection of kanamycin sulfate molecules with significant LOD around 0.85 fmol. By validating the excitation of toroidal dipoles in the mid-IR band, we qualitatively investigated the sensing performance of the tailored metasensor for several concentration of the targeted biological molecules. The high sensitivity and LOD of the detection mechanism were originated from the remarkable sensitivity of toroidal meta-atom to the minor variations in the dielectric permittivity of the media. We envision that the proposed toroidal metasensor technology possesses strong potential to be employed for practical and modern on-chip biosensing application.

**Methods**

**Simulation.** To accurately predict the plasmonic properties of the proposed toroidal mid-IR metamaterial, we investigated the excitation of the dipolar moment and its interference using finite-difference time-domain (FDTD) method (Lumerical 2019). In our simulations, to extract the plasmonic response of the tailored metastructure,[32-41] following parameters were used: The spatial cell sizes were set to 3 nm in all three axes, with 28 perfectly matched layers (PMLs) as absorptive the boundaries. The mid-IR light source was a plane wave with a simulation duration of 600 fs. The dielectric functions for the gold meta-atoms were taken from empirically reported constants by Ordal *et al*.[42] The current density profile in Supporting

Information file was calculated using the displacement current simulation module based on implementing the effective permittivity of the system ($\vec{D} = \varepsilon_{eff}\vec{E}$). Subsequently, the current density was solved in FDTD simulations via the following equation: $\vec{J} = -i\omega(\varepsilon_{ff} - \varepsilon_0)\vec{E}$.

**Sample fabrication.** The samples were fabricated on a 500 μm-thick quartz substrate using electron-beam lithography (EBL) technique. The wafers were developed by PMMA layer and later 3 nm of chromium was deposited on the PMMA layer to enhance conduction. The developed unit cells were patterned EBL system. Then an 70 nm-thick layer of gold was deposited using an e-beam evaporator with the vacuum pressure of $6.7 \times 10^{-7}$ Torr. Ultimately, the metallized specimens were immersed in acetone for approximately an hour for liftoff.

**Conflicts of interest**

There are no conflicts to declare.

**Figures**

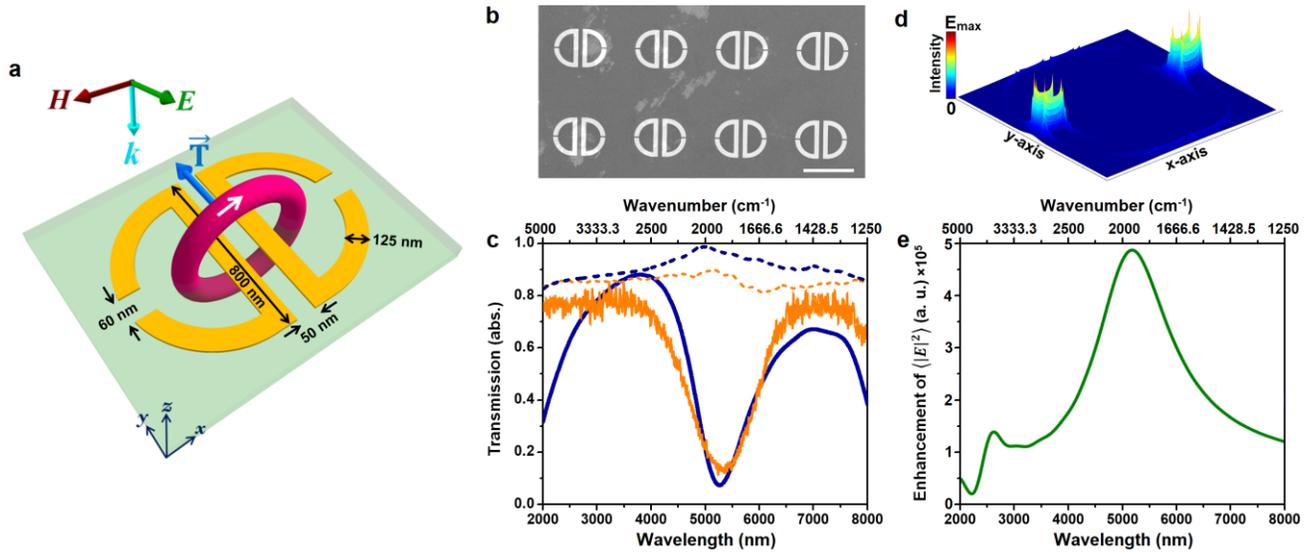

Figure 1. **The spectral response of the toroidal metasurface. a**) Schematic of the proposed mid-IR plasmonic toroidal meta-atom with the description to the judiciously defined geometries under longitudinal plane wave excitation. The circular spinning feature artificially shows the direction and formation of the head-to-tail charge-current configuration around the unit cell. **b**) SEM image of the fabricated metasurface. **c**) Experimentally (orange) and numerically (blue) obtained normalized amplitude transmission spectra for the excitation of toroidal moment under longitudinal (solid) and transverse (dashed) polarization excitations. **d**) 3D E-field enhancement maps for the excitation of toroidal dipole, showing the strong confinement of plasmons at the capacitive gap spots. **e**) E-field enhancement ($|E|^2$) as a function of incidence, validating significant intensity of the confined fields at the toroidal dipole wavelength.

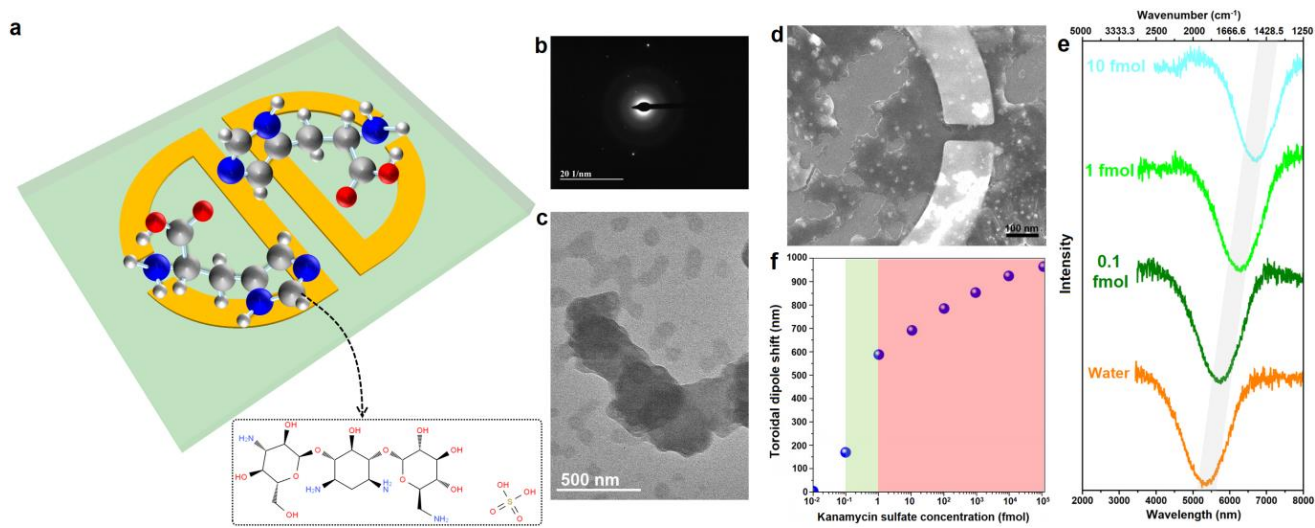

Figure 2. **Principle of biomolecule detection using toroidal plasmonic metamaterial. a**) Schematic of the proposed mid-IR plasmonic toroidal meta-atom with the presence of kanamycin sulfate molecules deposited on the surface. The inset shows the molecular structure for the targeted kanamycin sulfate molecules. **b**) dark- and **c**) bright-field images for the diffraction pattern and transmission electron microscope (TEM) images of molecules. **d**) SEM image of an area of meta-molecule with the presence and accumulation of kanamycin sulfate molecules at the capacitive opening of the metastructure. **e**) The experimentally measured transmission spectra for various concentration of targeted antibiotic molecules. This panel shows a continuous and monotonic red-shift in the position of the toroidal dipole by increasing the density of biological objects. **f**) Toroidal dipole position value as a function of the kanamycin sulfate molecules concentration. The shaded red region shows the starting point for the reaction of the toroidal mode to the presence of antibiotic molecules.

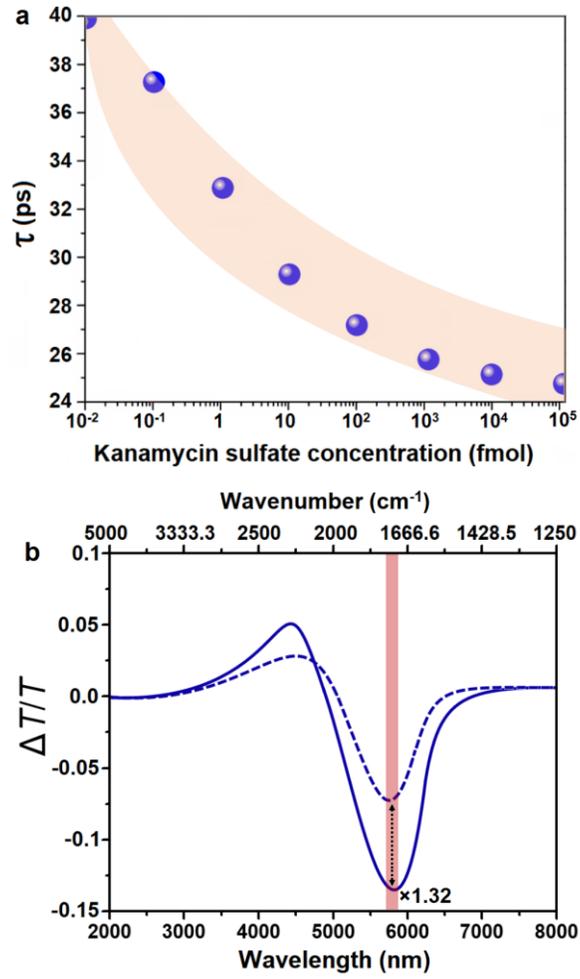

Figure 3. **a**) The lifetime of toroidal dipole as a function of antibiotic molecules concentration. **b**) Experimentally measured functional change of transmission spectra before (dashed) and after (solid) binding of kanamycin sulfate molecules to the meta-atoms.

**Supporting Information**

1. **Vectorial surface current density map**

The surface current density profile is computed using the displacement current ($\vec{J}$) simulation module based on implementing the effective permittivity of the developed mid-IR plasmonic system ($\vec{D} = \varepsilon_{eff}\vec{E}$). Therefore, the corresponding vectorial surface current density can be solved in FDTD simulations *via* the following equation: $\vec{J} = -i\omega(\varepsilon_{eff} - \varepsilon_0)\vec{E}$. The mismatch between the induced magnetic (**m**) fields in proximal pixels verifies the excitation of toroidal dipole in the unit cell at the targeted wavelength ($\lambda \sim 5300$ nm), as shown in Figure S1. This panel also demonstrates the direction of the toroidal moment and enhancement of surface current at the center of the unit cell.

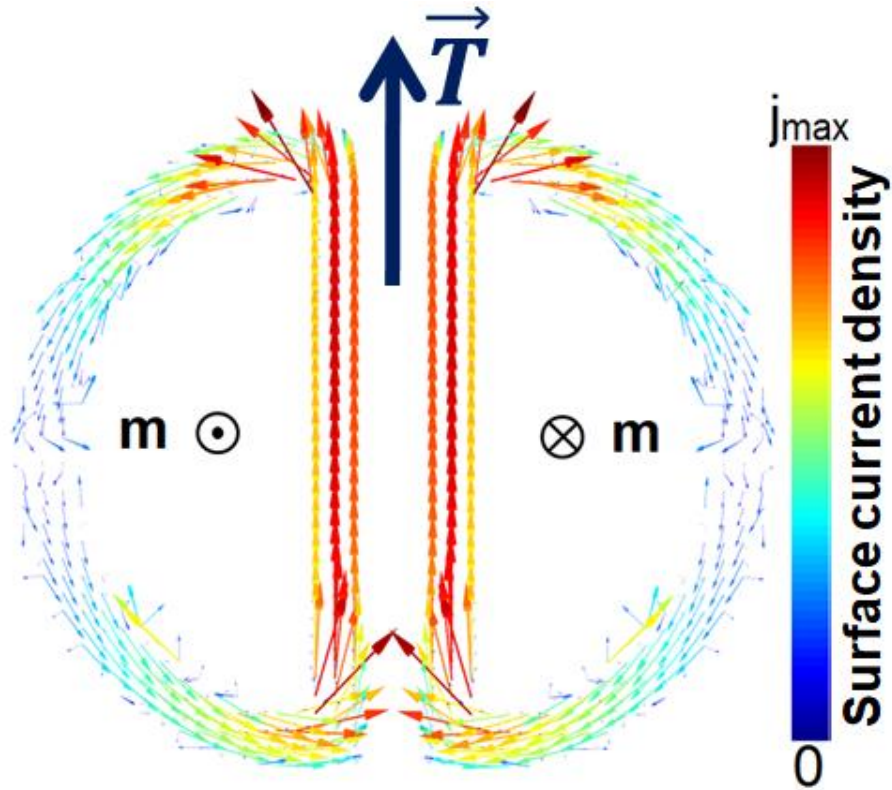

**Figure S1.** Vectorial surface current density map across the unit cell at the toroidal resonance wavelength.

## 2. Multipolar decomposition (Radiated power from multipoles)

In Figure S2, we calculated and explicitly exhibited the excitation of toroidal spectral feature by showing the strength of multipole excitations for the power of the scattered multipoles emitting from the plasmonic scatterer. In this mode analysis profile, the main contributors to the unit cell spectral response are the electric dipole (p), in conjunction with a superposition of well-matched toroidal dipolar (T), and magnetic quadrupolar ($Q^{(m)}$) moments.

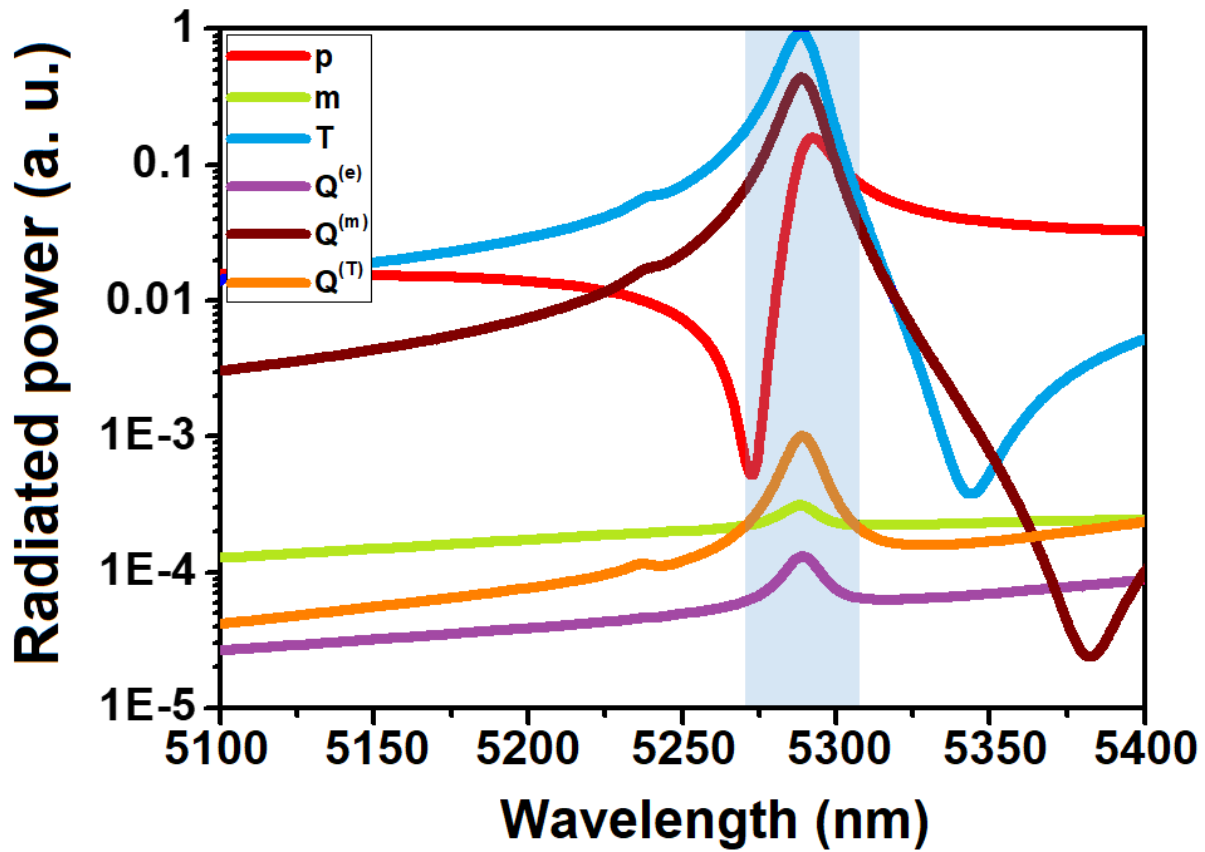

**Figure S2.** Calculated radiated power for individual electromagnetic multipoles excited in meta-atom by the incident beam illumination.

## 3. Vectorial magnetic-field

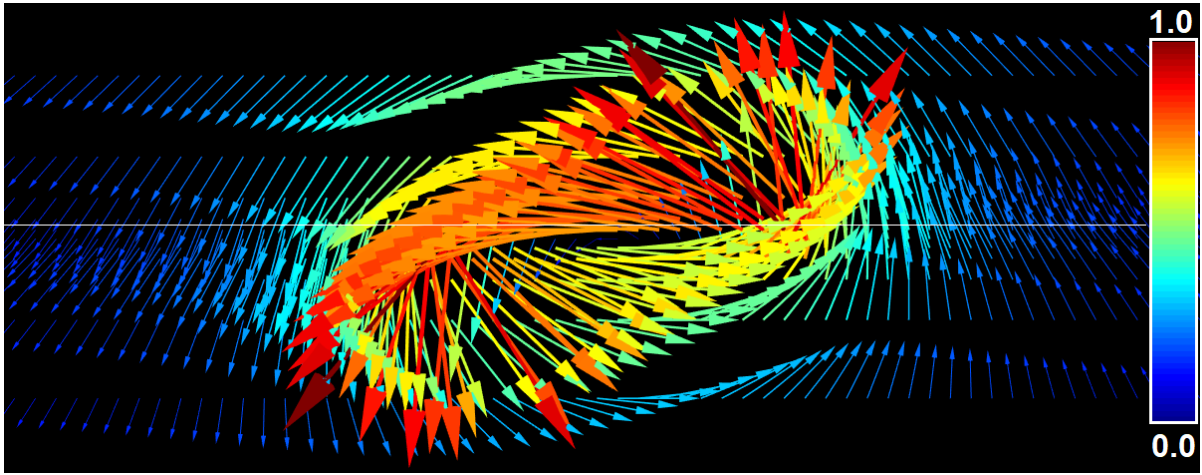

**Figure S3.** Vectorial cross-sectional (*xz*-plane) magnetic-field for the toroidal dipole mode.

## 4. Sensing the variations in the refractive index of the media

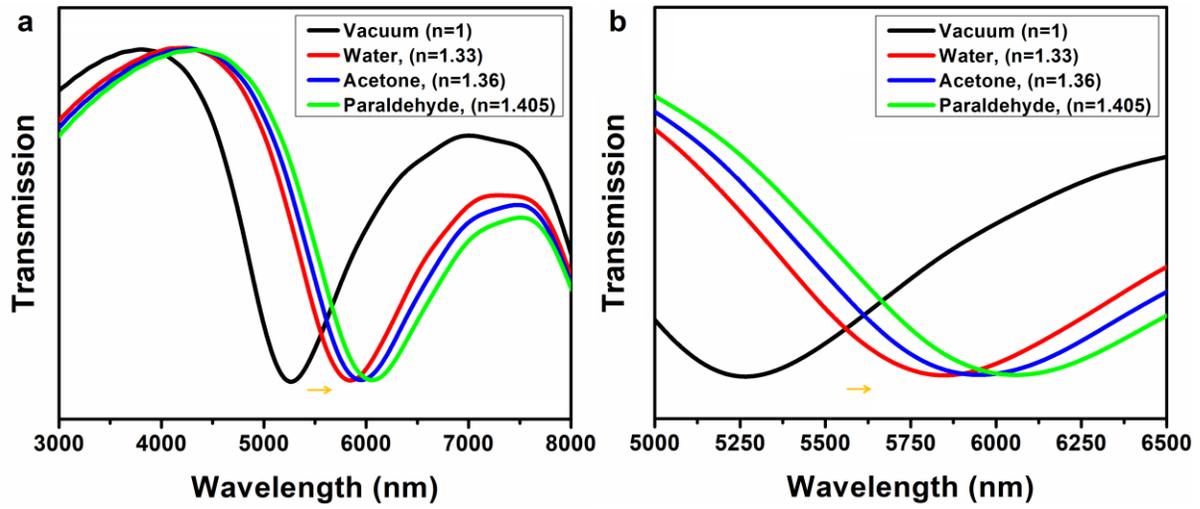

**Figure S4.** a) The transmission spectra for varying the refractive index of the medium around the chip. b) The magnified panel for the shift in the position of the toroidal dipole due to refractive index variations.